\documentclass[secnumarabic,amssymb, amsmath, amsfonts , nofootinbib,showpacs, aps, prl,superscriptaddress,preprint]{revtex4}
\usepackage{amsfonts}
\usepackage{amsmath}
\usepackage{amssymb}
\usepackage{graphicx}%
\usepackage{bm}%
\usepackage{graphicx}
\begin{document}
\input{epsf}
\def\Fig#1{Figure \ref{#1}}
\def\Eq#1{Eq.~(\ref{#1})}
\def\Table#1{Table~\ref{#1}}

\title{The driving factors of electro-convective instability in concentration polarization}

\author{Isaak Rubinstein}\email{robinst@bgu.ac.il.}\affiliation{Jacob Blaustein Institutes for Desert
Research, Ben-Gurion University of the Negev, Sede Boqer Campus
8499000, Israel}
\author{Boris Zaltzman }\email{boris@bgu.ac.il.}\affiliation{Jacob Blaustein Institutes for Desert Research, Ben-Gurion
University of the Negev, Sede Boqer Campus 8499000, Israel}


\begin{abstract}
Ionic current in a binary electrolyte passing through a charge-selective interface (electrode, ion exchange membrane, micro-nano-channel junction) is a basic element of many electrochemical engineering and micro-fluidic processes, such as electrodeposition, electrodialysis and protein pre-concentration. Such current passage is diffusion-limited in the sense that it induces a decrease of electrolyte concentration towards the interface (concentration polarization) whose expression is the saturation of current upon increasing voltage at some value -- the limiting current. Upon a further increase of voltage, this saturation is followed by a relatively rapid current increase -- over-limiting conductance regime. It is commonly accepted that in open systems over-limiting conductance is mediated by a micro-scale vortical flow which spontaneously develops as a result of electro-convective instability of quiescent concentration polarization near the limiting current. Electro-convection is a flow driven by the electric force acting either upon the space charge of the interfacial electric double layer (electro-osmosis) or the residual space charge of the quasi-electroneutral bulk (bulk electro-convection). There are two types of electro-osmosis, the equilibrium and the non-equilibrium one, the former relating to the action of the tangential electric field upon the space charge of the electric double layer, and the latter pertaining to the similar action upon the extended space charge which forms next to the electric double layer near the limiting current. For a perfectly charge-selective interface, concentration polarization under the equilibrium electro-osmotic slip condition is stable, and so it is with respect to bulk electro-convection, as opposed to non-equilibrium electro-osmosis which may cause instability. For this reason until recently, the electro-convective instability in concentration polarization was attributed to this latter mechanism. Lately, it was shown that imperfect charge-selectivity of the interface makes equilibrium instability possible, driven by either equilibrium electro-osmosis or bulk electro-convection, or both. In this paper we identify and analyze the major surface and bulk factors affecting the electro-convective instability. These factors, some known previously under the names of diffusio-osmosis, electro-osmosis or bulk electro-convection, and some newly identified in this paper, are manifestations of the electric force and pressure gradient, balanced by the viscous force acting in various locations in solution. The contribution of these factors to hydrodynamic stability in concentration polarization is analyzed for a varying perm-selectivity of the interface.
\end{abstract}

\maketitle \section {Introduction} DC ionic current in a binary electrolyte passing through a perm-selective interface (electrode, ion exchange membrane, micro-nano-channel junction) is a basic element of many electrochemical engineering and micro-fluidic processes, such as electrodeposition, electrodialysis and protein pre-concentration, \cite{1}--\cite{21}. Such current passage is diffusion limited in the sense that it induces a decrease of electrolyte concentration towards the interface, the phenomenon known as concentration polarization (CP). A common expression of it is a characteristic voltage-current curve with a segment in which the current nearly saturates at some plateau value, the limiting current, corresponding to the nearly vanishing interface concentration.  This segment of the voltage-current curve is usually followed by a region of a relatively rapid increase of electric current with voltage -- the so-called over-limiting conductance (OLC) regime. The mechanism of OLC remained unexplained for a long time. Only recently was it shown that in open systems OLC is due to the destruction of the diffusion layer by a micro-scale vortical flow. This flow spontaneously develops as a result of a hydrodynamic instability of CP near the limiting current and provides an additional ionic transport mechanism yielding OLC, \cite{16}, \cite{22}--\cite{28}. This flow may be driven by the electric force acting both upon the space charge of a nanometers-thick interfacial electric double layer (EDL) and the residual space charge of the stoichiometrically electroneutral bulk. A slip-like fluid flow induced by the former is known as electroosmosis (EO), whereas the flow induced by the latter is referred to as bulk electro-convection (EC). There are two regimes of EO that correspond to different states of the EDL and are controlled by the non-equilibrium voltage drop (overvoltage) across it, \cite{24}, \cite{29}--\cite{33}: equilibrium EO, \cite{29}--\cite{31}, and non-equilibrium EO, or EO of the Second Kind, \cite{16}, \cite{32}, \cite{33}. While both regimes result from the action of a tangential electric field upon the space charge of the EDL, the former relates to the charge of the equilibrium EDL, whereas the latter relates to the extended space charge (ESC) of the non-equilibrium EDL which develops in the course of CP near the limiting current, \cite{34}--\cite{40}.

The theory of equilibrium EO at a perm-selective interface was developed by Dukhin and Derjaguin [30]. An essential component of this theory is accounting for polarization of the EDL by the applied tangential electric field, resulting in a lateral pressure drop in the EDL, owing to the lateral variation of the Maxwell stress. This yields for equilibrium EO slip velocity, instead of the common Helmholtz-Smoluchowski formula $u=-\zeta E$, the expression \cite{30}, \cite{33}
\begin{equation}
u=\zeta\left({\varphi}_x+\frac{{c}_x}{{c}}\right)+\frac{{c}_x}{{c}}\left(4\ln2-4\ln\left(e^{\zeta/2}+1\right)\right).\label{1}
\end{equation}
Here $\varphi$ is the dimensionless electric potential (scaled with the thermal potential, $kT/e$), $c$ is the dimensionless electrolyte concentration (scaled with some typical concentration $c_0$), $x$ is the dimensionless tangential coordinate (scaled with some typical macroscopic length, e.g. the membrane width), and $\zeta$ is the dimensionless electric potential drop between the interface and the outer edge of the EDL, and $E=-\varphi_x$ is the dimensionless electric field. The peculiarity of (\ref{1}) is that, for an ideally perm-selective cation exchange membrane maintained at a fixed electric potential, the electrochemical potential of counterions in the membrane, $\ln c+\varphi=\textrm{const.}$, is constant, and so it is, in equilibrium conditions, at the outer edge of the EDL. In other words, $\partial {c}/\partial x=-c\partial {\varphi}/\partial x$, and for $\zeta\to-\infty$, equation (\ref{1}) yields
\begin{equation}
{u}=4\ln2 E\label{2}
\end{equation}
Hydrodynamic stability of the quiescent CP with a limiting equilibrium EO slip condition (\ref{2}) was studied by Zholkovskij et al., \cite{41}, who found that 1D CP was stable. So it was concluded that with a perfectly perm-selective interface no bulk EC instability (ECI) was feasible for a low molecular electrolyte, \cite{33}. In brief, the physical reason for this is that for an ideal interface, the stabilizing Donnan contribution to the electric potential perturbation, resulting from the concentration perturbation by the flow, dominates the corresponding destabilizing Ohmic contribution. Recognizing this balance has motivated the reexamination of the role of perfect charge selectivity of the interface, \cite{42}. (For a detailed discussion of bulk EC versus equilibrium EO and the extent to which the two are equivalent, see \cite{42}.) On the other hand, it was shown that the non-equilibrium slip related to the ESC did yield instability, \cite{24}, \cite{33}, \cite{43}. This was the reason why since its prediction in 1999, \cite{43}, till now hydrodynamic instability in CP was attributed to non-equilibrium EO and was so studied, \cite{16}, \cite{18}, \cite{19}, \cite{43}--\cite{47}.

In \cite{42} we showed that any deviations from constancy of the electrochemical potential of counter-ions at the outer edge of EDL makes equilibrium instability possible, driven by either equilibrium EO or bulk EC, or both. Non-constancy of the counter-ionic electrochemical potential may result either from non-ideal perm-selectivity of the interface, addressed in \cite{42}, or from a finite rate of electrode reactions (e.g., in cathodic deposition).

Thus, depending on the system, in particular, on the perm-selectivity of the charge selective element in it, the ECI in CP at a non-perfect charge selective interface may be affected, either driven or inhibited, by several factors, such as equilibrium or non-equilibrium EO, bulk EC, equilibrium and non-equilibrium diffusio-osmosis, etc. Some of these factors, e.g. equilibrium EO and diffusio-osmosis and bulk EC, are accounted for in the locally quasi-electroneutral models. Other factors, e.g. non-equilibrium EO and diffusio-osmosis, affect ECI only when we take into account the possibility of the violation of local electro-neutrality in the bulk, that is, the formation of ESC. Also, the effect of these factors on the hydrodynamic stability may vary depending on the perm-selectivity of the system and the voltage regime in it. The previous studies addressed the limiting cases in which some of these factors either were absent or affected the stability in a predetermined manner. Thus, in a perfectly perm-selective system the destabilization resulted solely from the ESC, whereas in the recent study of the equilibrium instability all ESC related effects were disregarded completely.

In this study we embark at identifying and analyzing the major 'surface' and 'bulk' factors driving the hydrodynamic instability in concentration polarization at a non-perfect charge selective interface. In brief, these factors are four:
\newline 1. Tangential variation of the electrolyte concentration, equivalent to the tangential variation of conductivity, induced space charge, etc. On the whole, this factor may be identified as generalized diffusio-osmosis.
\newline 2. Tangential regular (applied) electric field acting upon the space charge of quasi-equilibrium EDL or the residual space charge of the quasi-electroneutral bulk. This factor may be identified as the generalized induced charge EO (ICEO, \cite{31}), and its bulk electroconvection analog.
\newline 3. Tangential variation of the counter-ionic transport number in the system. This factor has to do with the tangential variation of the logarithmic potential drop normal to the interface, forming in the bulk solution in the vicinity of the interface in the course of CP. This drop, which we term 'singular' for brevity and its related tangential electric field, becomes a major factor at and above the limiting current. This factor may be viewed as a singular bulk analog of the induced charge EO (SB-ICEO).
\newline 4. Tangential variation of the total normal ionic mass flux. This factor, which may be vaguely termed fricto-osmosis, is entirely related to the extended space charge of the non-equilibrium EDL forming at and above the limiting current. This factor is a direct generalization of the non-equilibrium EO or EO of the second kind, analyzed previously for ideally perm-selective interfaces, \cite{24}, \cite{33}, \cite{43}.

\noindent The second and the third factors are related to the regular (factor 2 - for under-limiting current regime) and singular (factor 3 - for limiting and over-limiting regimes) components of the tangential electric field.

All four factors are merely various expressions of the electric force and pressure gradient in balance with the viscous force acting in various locations in solution, whereas their related flows bear various names, such as diffusio-osmosis, electro-osmosis, bulk electroconvection, etc.

As in our previous studies for brevity we assume equal ionic diffusivities the same for the membrane and the solution, and so a constant dielectric permeability (equal for the membrane and for the solution). For the effects related to the variations of this latter induced by temperature, concentration or the electric field itself, the reader is referred to \cite{48} and references therein.
Below we show  that hydrodynamic instability in concentration polarization  occurs near the limiting current and is driven by the factors 3 and 4. For low electrolyte concentration (high perm-selectivity), instability is mediated by fricto-osmosis, (factor 4), related to the extended space charge, whereas upon the increase of electrolyte concentration (decrease of perm-selectivity) ICEO and SB-ICEO (factors 2 and 3) take over.

\section {Driving factors of hydrodynamic instability in CP.} Following \cite{42}, we begin by formulating a three-layer model for a membrane flanked by two concentration polarized diffusion layers whose stability under no-slip (bulk electro-convection-no-slip setup) and equilibrium slip condition (\ref{1}) (bulk electro-convection-EO setup) we analyze.

\begin{figure}[!htb]
\includegraphics[width=\columnwidth,keepaspectratio=true]{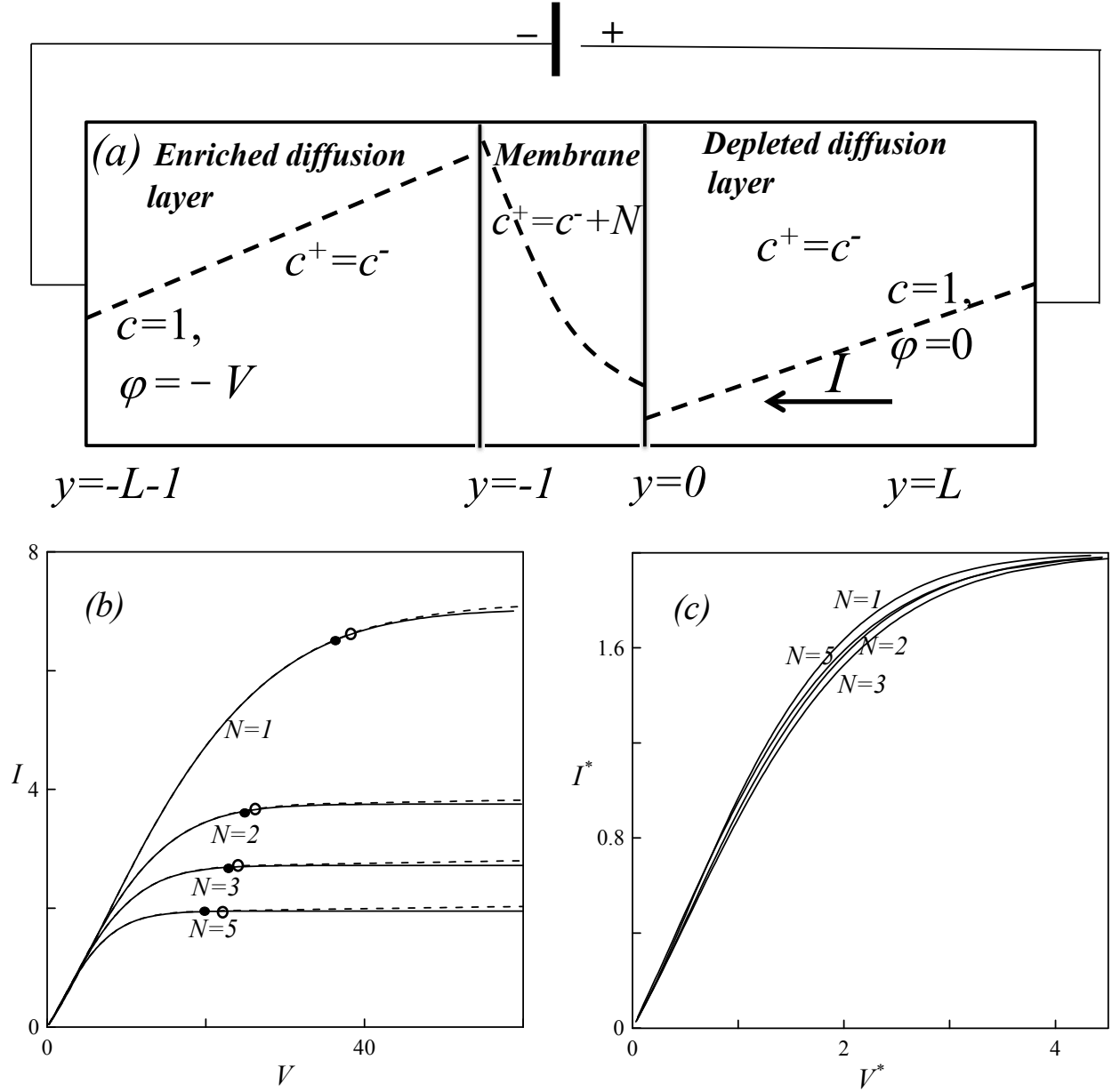}
\caption{(a) Scheme of three-layer setup, dashed lines -- schematic plots of the average ionic concentration, $C(y)$; (b) Unscaled voltage-current-dependence; Solid line stands for full setup, $\varepsilon=10^{-4}$ and dashed line stands for the limiting electroneutral setup. The black dots and circles mark the onset of instability for the full and the limiting electroneutral setups, correspondingly; (c) Same plots for scaled voltage-current-dependencies.}
\end{figure}

Let us consider an infinite 2D cation-exchange membrane, $-\infty<x<\infty,\ -1<y<0$, flanked by two diffusion layers, $-\infty<x<\infty,\ -L-1<y<-1$ and $-\infty<x<\infty,\ 0<y<L$, of a univalent electrolyte with equal fixed concentrations and a given drop of electric potential maintained in the reservoirs at the outer boundary of diffusion layers (see Fig.1b). This three-layer system is modeled by the following boundary-value problem non-dimensionalized in a natural manner, \cite{42},
\begin{align}
&\frac{\partial c_\pm}{\partial t}=-\mathbf{\nabla}\cdot{\mathbf{j_\pm}};\label{3}\\
&\mathbf{j}_{\pm}=-c_{\pm}\mathbf{\nabla}\mu_{\pm}+\textrm{Pe} \mathbf{v}c_{\pm}, \ \mu_{\pm}=\ln{c_{\pm}}\pm \varphi;\label{4}\\
&\varepsilon^2\nabla^2\varphi=N(x)+c_--c_+.\label{5}\
\end{align}
Here $c_\pm$ is the concentration of positive and negative ions, $\textrm{Pe}$ is the material Peclet number, $\varepsilon$ is the dimensionless Debye length and
\begin{align}
&N(x)=0,\  -L-1<y<-1, \ 0<y<L;\label{5a}\\
&N(x)=N,\ -1<y<0,\label{5b}\
\end{align}
where $N$ is the dimensionless fixed charge density in the membrane (scaled by the product of positive elementary electric charge and the dimensional reservoir concentration), \cite{14}, \cite{16}.
We neglect the fluid flow in the enriched diffusion layer and in the membrane: $\mathbf{v}=u\mathbf{i}+w\mathbf{j}\equiv 0,\ -L-1<y<0$; and determine it in the depleted diffusion layer from the Stokes-continuity equations:
\begin{equation}
\mathbf{\nabla}^2 \mathbf{v}-\mathbf{\nabla}p+\mathbf{\nabla}^2\varphi\mathbf{\nabla}\varphi=0,\ \mathbf{\nabla}\cdot\mathbf{v}=0.\label{6}
\end{equation}
At the outer boundary of the depleted diffusion layer we apply the reservoir conditions for velocity, $u_y(x,L)=w(x,L)=0$, along with prescribing the concentration and the electric potential at the outer edges of both diffusion layers:
\begin{equation}
c_+=c_-|_{y=-L-1,L}=1,\ \varphi|_{y=-L-1}=-V,\ \varphi|_{y=L}=0.\label{8}
\end{equation}
We complete the formulation by prescribing continuity of the ionic concentrations, $c_{\pm}$, electric potential, $\varphi$, together with their gradients through the membrane--solution interfaces, $y=-1,0$ and the non-slip condition at the membrane--depleted diffusion layer interface, $y=0$. The main control parameters are  the dimensionless voltage $V$, the dimensionless width of the diffusion layers, $L$, and $N$. The latter is the measure of membrane charge selectivity, a perfect membrane corresponding to $N>>1$. For these conditions, the counterion concentration in the membrane equals $N$, which, combined with a fixed electric potential, amounts to fixing the electrochemical potential of counterions employed in the previous one-layer models, \cite{24}, \cite{33}, \cite{43}. In the three-layer model, reducing $N$ from infinity (perfect membrane) to a practical range, $N>1$, amounts to allowing for lateral variations of the electrochemical potential of counterions in the membrane. The flow in the enriched compartment and the possible EO flow across the membrane are disregarded for simplicity, recognizing that hydrodynamic instability in concentration polarization is entirely due to large electric fields in the depleted diffusion layer.

In Ref.\cite{42} we addressed the quasi-equilibrium-electroneutrality asymptotic limit, $\varepsilon=0$, of the problem (\ref{3}--\ref{8}). In this limit
\newline A) The Poisson eq. (\ref{5}) is reduced to electroneutrality conditions in the enriched, $-L-1<y<-1$, and in the depleted, $0<y<L$ diffusion layers and the membrane, $-1<y<0$, which read, respectively:
\begin{align}
&c_+=c_-=C,\ -L-1<y<-1, \ 0<y<L;\label{9}\\
&c_+=c_-+N=C+\frac{N}{2},\ -1<y<0.\label{10}\
\end{align}
Here $C=(c_++c_-)/2$ is the average ionic concentration.
\newline B) The continuity of the ionic electrochemical potentials, $\mu_{\pm}$, and normal ionic fluxes holds across the membrane--solution interfaces, $y=-1,0$.
\newline C) The non-slip condition is replaced by the "outer" slip condition (\ref{1}) at the membrane--depleted diffusion layer interface, $y=0$.

The quiescent 1D steady-state solution to the problem (\ref{1}), (\ref{3}), (\ref{4}), (\ref{6}), (\ref{8}-\ref{10}) has been computed analytically in terms of Lambert functions and its stability has been studied, \cite{42}. In Figs.1--3 we compare the results for full model problem  (\ref{3})--(\ref{8}) and its asymptotic limit (\ref{1}), (\ref{3}), (\ref{4}), (\ref{6}), (\ref{8}--\ref{10}). Thus in Fig.1b we present the computed voltage-current dependencies for various $N$ (current density $\mathbf{I}$, is defined as $\mathbf{I}=\mathbf{j}^+-\mathbf{j}^-$; in the figures below $I$ is the normal component of $\mathbf {I}$ averaged over the interface). We note that, whereas the voltage-current curves computed for different $N$ strongly differ due to the decrease of membrane perm-selectivity with the decrease of $N$ (Fig.1b), upon a suitable scaling, the scaled $I^*-V^*$ curves collapse. Here $I^*=I/I_0,\ V^*=V/V_0$, where $I_0$ is, e.g., one half of the limiting current, $I_{lim}$, and $V_0$ is the corresponding voltage.

\begin{figure}[!htb]
\includegraphics[width=\columnwidth,keepaspectratio=true]{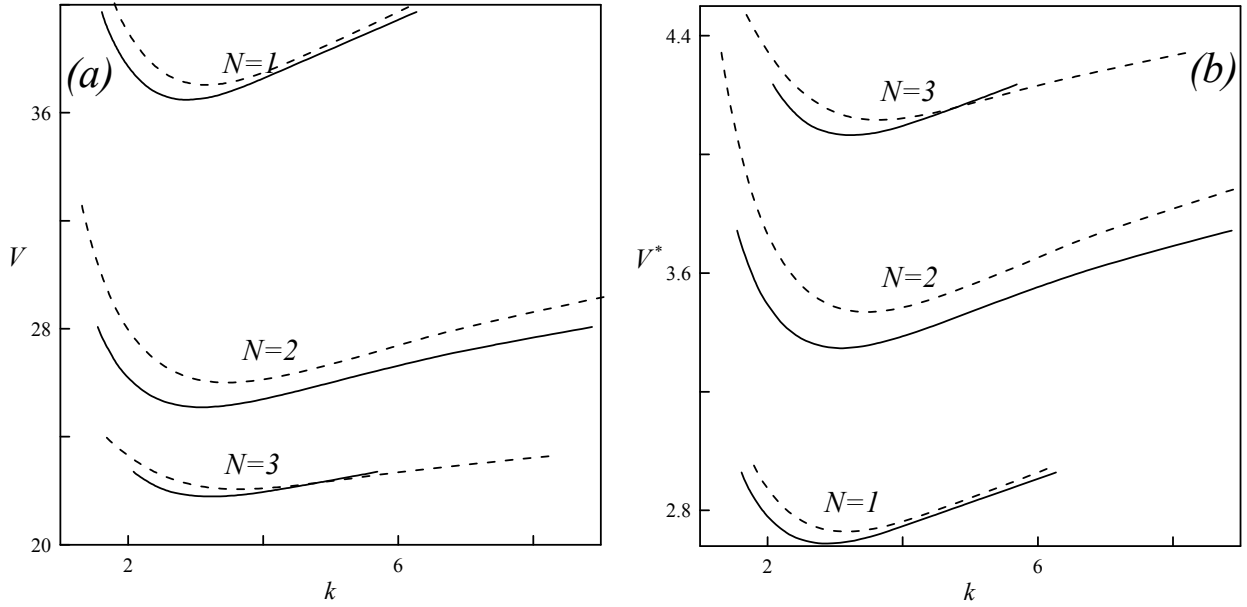}
\caption{(a) Neutral stability curves in unscaled voltage $V$--wave number $k$ plane (above the curve -- instability), $L=1$;  $N=1$, $N=2$,  $N=3$. Solid line stands for full setup, $\varepsilon=10^{-4}$ and dashed line stands for the limiting electroneutral setup; (b) Same for the scaled voltage $V^*$.}
\end{figure}

The results of the linear stability analysis of the quiescent 1D steady-state are presented in Fig.2,3. We note that whereas for the unscaled voltage the unstable portion of the $V$-$k$ plane shrinks exponentially upon the decrease of membrane perm-selectivity, for the scaled voltage the unstable portion of the voltage-wave number phase plane expands with the decreasing $N$. Accounting for the space charge results in slight decrease of the critical voltage. Also, we note that in either formulation the instability sets on for developed CP. This is illustrated through marking with dots the onsets of instability on the voltage-current curves in Fig.1b.

\begin{figure}[!htb]
\includegraphics[width=\columnwidth,keepaspectratio=true]{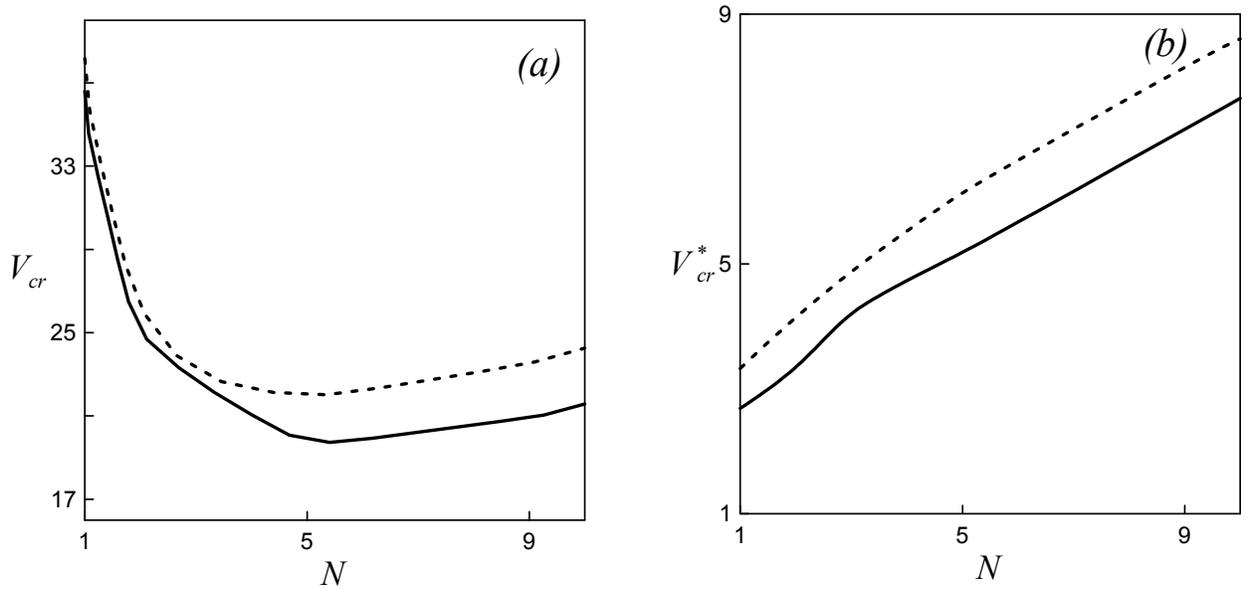}
\caption{(a) Unscaled critical voltage for full EC setup (solid line) and limiting electroneutral setup (dashed line) $V_{cr}$, versus $N$ for $L=1$ and $\varepsilon=10^{-4}$. (b) Same for scaled critical voltage $V^*_{cr}$.}
\end{figure}

In order to identify the major factors driving the hydrodynamic instability in CP at non-perfect charge selective interface, we employ the approach introduced in \cite{33} whose motivation is as follows. With the development of the ESC near the limiting current, the straightforward matched asymptotic expansion in the EDL analysis breaks down and the continuity of the ionic electrochemical potentials across the depleted interface fails \cite{33}. Nevertheless, even in these conditions for a planar interface, whenever the lateral length scale is long compared to that normal to the interface, there remains some fluid layer (extended boundary layer, EBL) through which the ion-transfer is essentially one-dimensional. This layer imbeds both the quasi-equilibrium EDL and the ESC. It has been shown in \cite{33} that the EBL scales with $\varepsilon$ as ${1}/{|\ln\varepsilon|}$, that is for reasonable voltages and realistically small $\varepsilon$ , it is much thicker than the EDL of either equilibrium or non-equilibrium (including the ESC) kind. Following \cite{33} we are about to analyze the one-dimensional transport equations in this EBL using the characteristics of the quasi-electroneutral bulk (QEB) as the boundary conditions. From this analysis, a boundary condition for the velocity at the outer edge of the EBL (the 'EO slip' velocity) will result.

\subsection{One-dimensional analysis in the EBL  $\quad 0<y<O\left(  {1}/%
{|\ln\varepsilon|}\right)$}
We consider the EBL at a non-perfect cation-exchange membrane $\left( 0<y<O\left[{1}/{|\ln\varepsilon|}\right] \right)  $, assuming that the right edge of the EBL lies in the quasi-electroneutral bulk, QEB. (Analysis of ESC at a non-perfect interface similar to the one to follow was recently carried out in \cite{49}.)

\textbf{Equations}:%
\refstepcounter{equation}
$$
   \frac{dc^{+}}{dy}+c^{+}\frac{d\varphi}{dy}    =-J^+,\quad
\frac{dc^{-}}{dy}-c^{-}\frac{d\varphi}{dy}    =-J^-,\quad
\varepsilon^{2}\frac{d^{2}\varphi}{dy^{2}}    =c^{-}-  c^{+}.
  \eqno{(\theequation{\mathit{a}-\mathit{c}})}\label{1.3}
 $$
Here: $J^+(x,t)\overset{\mathrm{def}}{=}-\left(\overline{c}_{y}(x,0,t)+\overline{c} \overline{\varphi}_y(x,0,t)\right)$, $J^-(x,t)\overset{\mathrm{def}}{=}-\left(\overline{c}_{y}(x,0,t)-\overline{c} \overline{\varphi}_y(x,0,t)\right)$ are, respectively, the boundary values of the cationic and anionic fluxes in QEB, where $\overline{c}$ and $\overline\varphi$ are the electrolyte concentration and electric potential in QEB.
Equations (\ref{1.3}\textit{a--c}) may be rewritten as follows
\begin{equation}
\varepsilon\frac{dc^{+}}{dy}=Ec^{+}-\varepsilon J^+,\label{1.10}
\end{equation}
\begin{equation}
\varepsilon\frac{dc^{-}}{dy}=-Ec^{-}-\varepsilon J^-,\label{1.11}
\end{equation}
\begin{equation}
\varepsilon\frac{dE}{dy}  =c^{+}-c^{-},\label{1.12}
\end{equation}
where
\begin{equation}
E=-\varepsilon\frac{d\varphi}{dy}. \label{1.13}%
\end{equation}
By substitution (\ref{1.11},\ref{1.12}) into (\ref{1.13}) and
integration of the resulting equation, we obtain%
\begin{equation}
c^{+}+c^{-}=\frac{1}{2}E^{2}-J\left(
y-y_{0}\right). \label{1.14}%
\end{equation}
Here $y_{0}$ is an integration constant and $J=J^++J^-$ is the salt flux.
By substituting (\ref{1.14}) into (\ref{1.10}, \ref{1.11}) we obtain the
following inhomogeneous Painleve equation of the second kind for $E$%
\begin{equation}
\varepsilon^{2}\frac{d^{2}E}{dy^{2}}=\frac{1}{2}E^{3}-J\left(  y-y_{0}\right)
E-\varepsilon I. \label{1.16}%
\end{equation}
Here $I=J^+-J^-$ is the electric current density. Considering (\ref{1.14}) in the electroneutral part of
EBL and keeping the leading terms, we conclude that
\begin{equation}
c^{+}=c^{-}=\overline{c}=-\frac{J}{2}\left(  y-y_{0}\right)=-\frac{J}{2}y+\overline{c}(x,0,t)\text{ in
EBL}\cap\text{QEB}. \label{1.17}%
\end{equation}
Thus, $y_{0}$ is the root of the linear extrapolation of the outer
(QEB) ionic concentration profile near the interface. Integrating (\ref{1.16})  in the electroneutral part of
EBL and keeping the leading order terms, we conclude that
\begin{equation}
\varphi (x,y,t)=k\ln(y-y_0)+\Phi(x,t)=k\ln\frac{2\overline{c}}{-J}+\Phi(x,t)=\overline{\varphi}\text{
in EBL}\cap\text{QEB}. \label{1.17b}%
\end{equation}
Here the flux ratio, $k\overset{\mathrm{def}}{=}I/J$, is related to the counter-ionic transport number, $\eta\overset{\mathrm{def}}{=}J^+/I=\frac{1}{2}\frac{1-k}{1+k}$. Another constant in EBL $\Phi(x,t)\overset{\mathrm{def}}{=}\phi(x,0,t)$ is the boundary value of the regular component of the electric potential in QEB (to be distinguished from the logarithmic component singular at the limiting current), defined as
\begin{equation}
\phi(x,y,t)\overset{\mathrm{def}}{=}\overline{\varphi}(x,y,t)-k\ln\frac{2\overline{c}}{-J}.
\label{1.17d}%
\end{equation}
The boundary values of the electroneutral electrolyte concentration $C(x,t)\overset{\mathrm{def}}{=}\overline{c}(x,0,t)$ and of the regular component of the electric potential,  $\Phi(x,t)$, the salt flux $J$ and the ionic transport number $k$ are the main control parameters in our analysis. To complete the formulation we complement the asymptotic condition at the outer edge of the EBL
\begin{equation}
E=-k\frac{\varepsilon}{y-y_{0}}+...\quad\text{for} y-y_{0}>>\varepsilon^{{2}/{3}},
\label{1.18}%
\end{equation}
by a boundary condition at the membrane-solution interface $y=0$. To obtain this condition, we integrate the Poisson-Boltzmann equation across the quasi-equilibrium EDL in the membrane, $y=-O(\varepsilon)$. This yields:
\begin{align}
\frac{\varepsilon^2}{2}\left(\frac{d\varphi}{dy}\right)^2|_{y=0}=&c^-(x,0,t)+c^+(x,0,t)-\nonumber \\
&\left( c^-(x,0,t)e^{\overline{\varphi}_m(x,0-,t)-\varphi(x,0,t)}+c^+(x,0,t)e^{-\varphi(x,0,t)
-\overline{\varphi}_m(x,0-,t)}\right)-\nonumber\\
&N(\overline{\varphi}_m(x,0-,t)-\varphi(x,0,t)).\label{1.19}\
\end{align}
Here $\overline{\varphi}_m$ is the electric potential in the quasi-electroneutral part of the membrane. Applying the electroneutrality condition for the cation and anion concentrations at the outer edge of this layer, $y=-O(\varepsilon)$, we obtain
\begin{equation}
c^-(x,0,t)e^{\overline{\varphi}_m(x,0-,t)-\varphi(x,0,t)}+N=c^+(x,0,t)e^{\varphi(x,0,t)
-\overline{\varphi}_m(x,0-,t)},\label{1.20}
\end{equation}
or
\begin{equation}
\overline{\varphi}_m(x,0-,t)-\varphi(x,0,t)=-\ln\frac{N+\sqrt{N^2+4c^-(x,0,t)c^+(x,0,t)}}{2c^+(x,0,t)}.\label{1.21}
\end{equation}
Finally, through substituting (\ref{1.20}) in (\ref{1.19}), we obtain:
\begin{align}
\frac{E(x,0,t)^2}{2}=&c^-(x,0,t)+c^+(x,0,t)-\sqrt{N^2+4c^-(x,0,t)c^+(x,0,t)}+\nonumber\\
&N\ln\frac{N+\sqrt{N^2+4c^-(x,0,t)c^+(x,0,t)}}{2c^+(x,0,t)}.\label{1.22}\
\end{align}
We note that the physical meaning of Eqs.(\ref{1.19}), (\ref{1.22}) is the balance of the Maxwell stress and the ionic osmotic pressure accross the quasi-equilibrium EDL in the membrane. This type of balance is a common element of the theory of EDL, \cite{29}. Integrating (\ref{1.16}) and applying the boundary conditions (\ref{1.18}), (\ref{1.22}) we compute the electric field and electric potential as functions of the control parameters $C(x,t)$,  $\Phi(x,t)$, $J$ and $k$.

To find the tangential velocity we have to solve the 'inner' problem resulting from substituting the pressure evaluated from the force balance in the direction normal to the interface into the tangential force balance. This yields the equation
\begin{equation}
-\frac{1}{2}\frac{\partial}{\partial x}\left[  \left(
\frac{\partial\varphi
}{\partial y}\right)  ^{2}\right] +\frac{\partial\varphi}{\partial x}%
\frac{\partial^{2}\varphi}{\partial
y^{2}}+\frac{\partial^{2}u}{\partial y^{2}}  =0,\quad
u|_{y=0}=0,\label{1.23}
\end{equation}
whose solution is to be matched with the respective QEB solution. We note that, generally, in a 2D setup allowing for a stream function formulation, pressure as a variable is superfluous. Nevertheless, it is a convenient tool for derivation of slip-like conditions through a boundary layer analysis we employ here. In the EBL$\cap$QEB equation (\ref{1.23}) maybe simplified through the following arguments. In this region the representation (\ref{1.17}), (\ref{1.17b}) holds for the ionic concentrations and electric potential, implying that the only dependence of $\varphi$ on $y$ resides in the $\ln(y-y_0)$-term. This allows to rewrite the momentum balance (\ref{1.23}) in the following form valid in the EBL$\cap$QEB, $y_0<y<<1$,
\begin{equation}
\frac{\partial^{2}}{\partial y^{2}}\left(u(x,y,t)+kk_x\left[\frac{\ln^2(y-y_0)}{2}+2\ln(y-y_0)\right]+\Phi_x(x,t)k\ln(y-y_0)\right)=0,\label{1.24}
\end{equation}
Equation (\ref{1.24}) implies that the expression
\begin{equation}
\frac{\partial}{\partial y}\left(u(x,y,t)+kk_x\left[\frac{\ln^2(y-y_0)}{2}+2\ln(y-y_0)\right]+\Phi_x(x,t)k\ln(y-y_0)\right)\label{1.241}
\end{equation}
is constant in the layer $y_0<y<<1$. By approaching this layer from QEB we conclude that this is a finite constant. Variation of the logarithmic terms in (\ref{1.241}) is large (particularly so, when $y$ approaches $y_0$). Nevertheless, the variation of their combination with $u$ in the parenthesis in (\ref{1.241}) is bounded. Thus, to the leading order the expression
\begin{equation}
u(x,y,t)+kk_x\left[\frac{\ln^2(y-y_0)}{2}+2\ln(y-y_0)\right]+\Phi_x(x,t)k\ln(y-y_0)\label{1.242}
\end{equation}
is constant in the EBL$\cap$QEB layer, $y_0<y<<1$. By inspection of (\ref{1.242}) we observe that $u(x,y,t)$ varies strongly across the EBL$\cap$QEB as opposed to the combination (\ref{1.242}) (This is particularly true in the vicinity of $y_0$). It appears natural to identify the expression (\ref{1.242}) as that portion of $u$ which fully forms in EBL and, thus, is constant in the QEB. The logarithmic terms  in this expression stand for the contribution of the electric field in the QEB. The problem with this identification is that the logarithmic terms  are singular for $y=y_0$. To eliminate this singularity the asymptotically valid logarithmic approximation for the potential variation across the EBL$\cap$QEB has to be replaced by that for the potential proper:
\begin{equation}
\ln(y-y_0)=\frac{\varphi(x,y,t)-\Phi(x,t)}{k}. \label{1.243}
\end{equation}
We note that, although the potential varies strongly across the EBL encompassing $y_0$, it remains bounded for a non-vanishing $\varepsilon$. This yields the following definition for the portion of $u$ constant in the QEB (fully formed in the non-locally electro-neutral portion of the EBL) which we term the 'reduced' tangential velocity, $U$:
\begin{align}
U\overset{\mathrm{def}}{=}&u(x,y,t)+\Phi_x(x,t)(\varphi(x,y,t)-\varphi(x,0,t))+(\ln k)_x\Bigl[\frac{(\varphi(x,y,t)-\varphi(x,0,t))^2}{2}+\nonumber\\
&\Bigl.(\varphi(x,y,t)-\varphi(x,0,t))(\varphi(x,0,t)-\Phi(x,t)+2k)\Bigr].
\label{1.25}\
\end{align}

Substituting (\ref{1.25}) in (\ref{1.23}) yields for $U$ the following boundary-value problem
\begin{align}
\frac{\partial^{2}U}{\partial y^{2}}=&-\varphi_{yy}(\varphi_x-\Phi_x)+\frac{1}{2}\left[\varphi_y^2\right]_x+(\ln k)_x\left[\varphi_{yy}(\varphi-\Phi+2k)+\varphi_y^2\right] \text{ in EBL}, \label{1.26}\\
& U(x,0,t)=0,\label{1.27}\\
& U_y=O(1)\sim 0 \text{ in EBL}\cap\text{QEB}.\label{1.28}\
\end{align}

The electric potential $\varphi$ to be substituted in the r.h.s. of (\ref{1.26}) is found through solution of (\ref{1.16}), (\ref{1.18}), (\ref{1.22}) for given values of the control parameters $\Phi,\ C, \ J, k$. Thus, from (\ref{1.26}--\ref{1.28}) the driving factors for the tangential velocity are the lateral derivatives of these parameters. Although $\varphi$ and $U$ are nonlinear functions of $\Phi,\ C, \ J, k$, the the r.h.s. of (\ref{1.26}) is a linear superposition of their lateral derivatives and, thus:
\begin{equation}
U=A_{C}(\Phi,C, J, k)C_x+A_{J}(\Phi,C, J, k)J_x+A_{k}(\Phi,C, J, k)k_x,
\label{1.29}
\end{equation}
in EBL$\cap$QEB. Here the multipliers at the tangential derivatives of  $\Phi,\ C,\ k$ and $J$ are the numerically computed partial derivatives of $U$ with respect to these parameters. All these multipliers are constant in the QEB (fully formed in the EBL) as opposed to those in the corresponding representation  of $u(x,y,t)$:
\begin{align}
&u=(\varphi(x,0,t)-\varphi(x,y,t))\Phi_x+A_{C}C_x+A_{J}J_x+\nonumber\\
&\left(A_{k}-\frac{(\varphi(x,y,t)-\varphi(x,0,t))^2}{2k}-\frac{\varphi(x,y,t)-\varphi(x,0,t)}{k}(\varphi(x,y,t)-\Phi(x,t)+2k)\right)k_x.
\label{1.29a}\
\end{align}
According to (\ref{1.17b}) the factors at $\Phi_x$ and $k_x$ in (\ref{1.29a}) are dominated in QEB$\cap$EBL, $y_0<y<<1$, by the $-k\ln(y-y_0)$ and $k{\ln(y-y_0)^2}/2$ terms, correspondingly.
In order to reduce the number of control parameters, in our modeling we return to the quiescent 1D formulation of the three-layer model problem  (\ref{3}--\ref{8}) and solving it find $J,\ C,\ k, \Phi$ as a function of a single parameter, the overall potential drop $V$. In Fig.4 we illustrate the formation in the EBL of the multipliers in representation (\ref{1.29}) for $y_0<y<<1$ along with that of the reduced velocity $U$.

\newpage
\begin{figure}[!htb]
\includegraphics[width=\columnwidth,keepaspectratio=true]{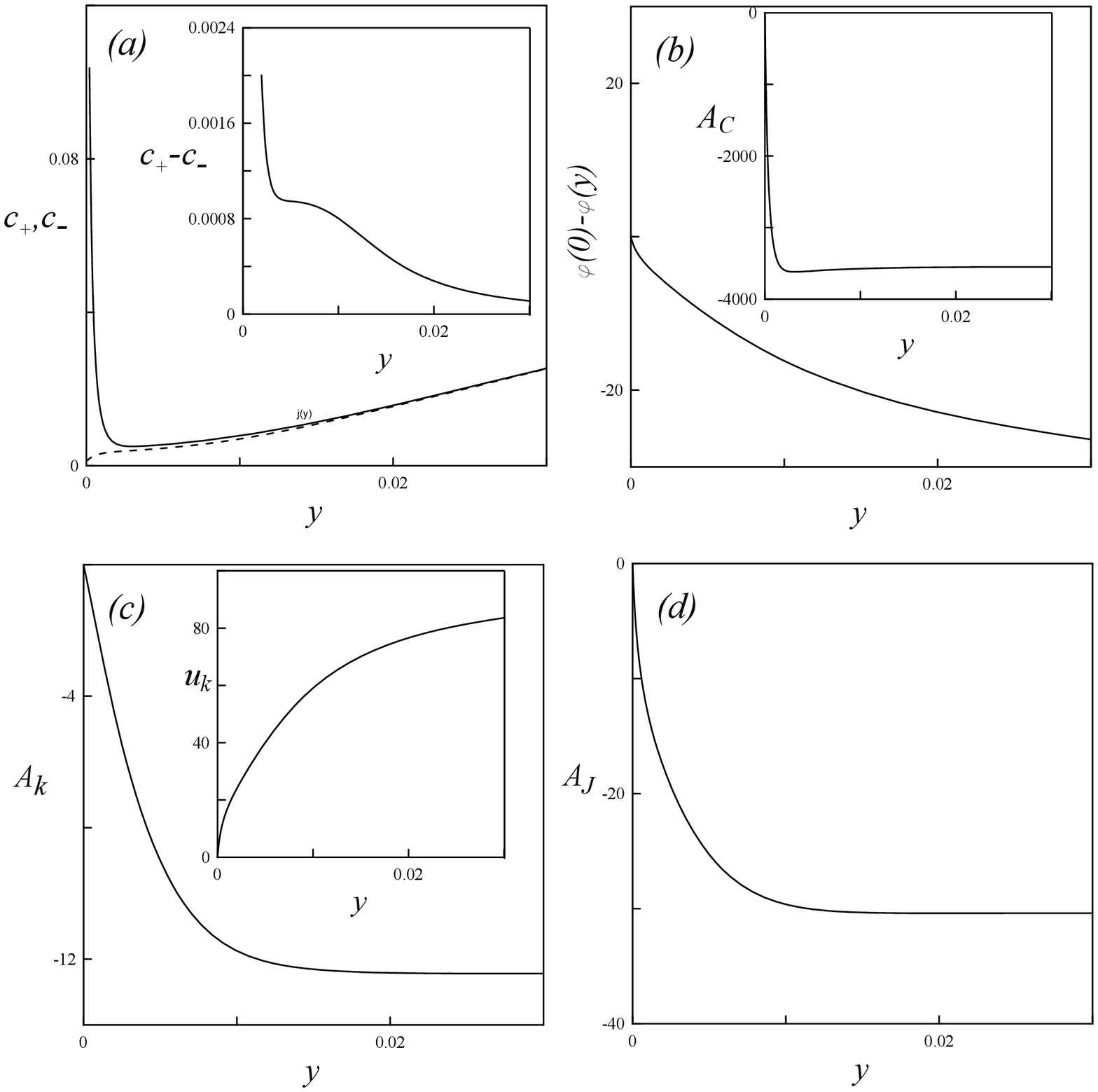}
\caption{Fine structure of the EBL and spatial development of the multipliers in the four driving factors, $\varepsilon=10^{-4},\ N=1,\ V=63.4,\ J=-2.01,\ \overline{c}=-J/2(x-1)+1,\ C=-0.005,\ y_0=0.005$
\newline (a) Concentration $c_+$ (solid line) and $c_-$ (dashed line); Inset: charge density $c_+-c_-$.
\newline (b) Velocity factor $u(y)/\Phi_x=\varphi(0)-\varphi(y)$ for $k_x=J_x=C_x=0$. Inset: Velocity factor $U(y)/C_x=u(y)/C_x$ for $\Phi_x=k_x=J_x=C_x=0$.
\newline (c) Velocity factor for reduced velocity $U(y)/k_x$, $\Phi_x=J_x=C_x=0$. Inset: same for full velocity $u(y)/k_x$, $\Phi_x=J_x=C_x=0$.
\newline (d)  Velocity factor $U(y)/J_x=u(y)/J_x$ for $\Phi_x=k_x=C_x=0$.}
\end{figure}

In Fig.5 we depict the dependence of these multipliers computed at $y=0.05$ on the scaled voltage $V^*$ for three values of $N=1/3, 1, 3$, corresponding to the weakly, moderately and highly charge-selective membrane.
\begin{figure}[!htb]
\includegraphics[width=\columnwidth,keepaspectratio=true]{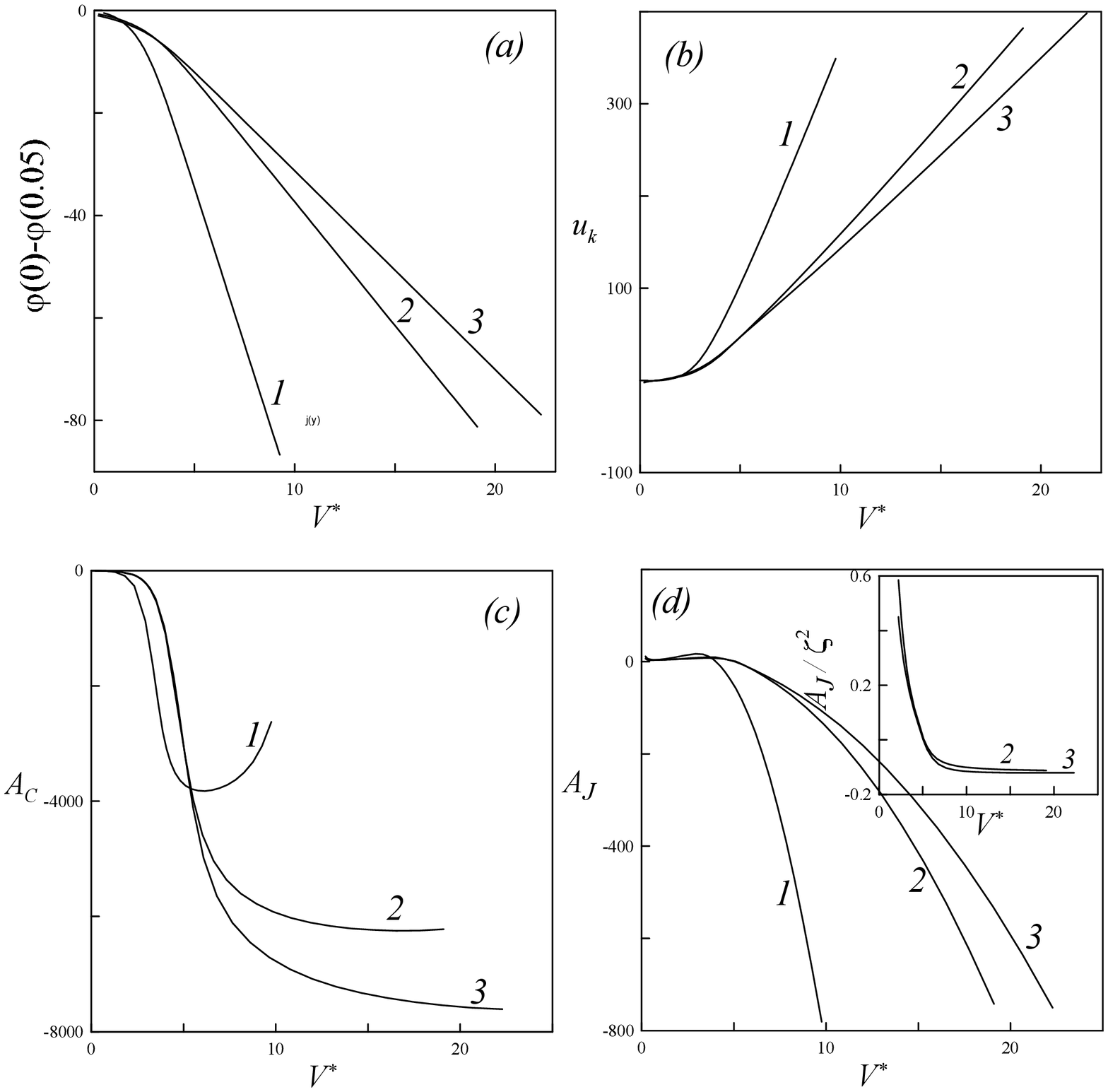}
\caption{Dependence of multipliers in the driving factors on the scaled voltage, $\varepsilon=10^{-4},\ N=1/3(1),\ 1(2),\ 3(3)$; $y=0.05$.
\newline (a) EO factor, $\varphi(0)-\varphi(0.005)$; (b) Bulk force factor $u_k=u(0.05)/k_x$ for $\Phi_x=J_x=C_x=0$; (c) Diffusio-osmotic factor, $A_C$;
(d)  Fricto-osmotic factor $A_J$. Inset: $A_J/\zeta^2,\ \zeta=\varphi(0)-\varphi(0.05)$.}
\end{figure}

To evaluate the effect of each factor in (\ref{1.29a}) on the stability of 1D conduction we consider the perturbation of the 1D quiescent steady-state by the accidental test vortex:
\begin{equation}
\bold{v_0}=u_0\bold{i}+v_0\bold{j}=-\frac{2y(y-1)(2y-1)}{iK}\mathbf{i}\exp(iKx)+\left(y(y-1)\right)^2\exp(iKx)\mathbf{j}\label{1.30}
\end{equation}
The correposnding time-dependent problem reads:
\begin{align}
&\frac{\partial c_\pm}{\partial t}=-\mathbf{\nabla}\cdot{\mathbf{j_\pm}};\label{1.31}\\
&\mathbf{j}_{\pm}=-c_{\pm}\mathbf{\nabla}\mu_{\pm}+\textrm{Pe} (\mathbf{v}+\mathbf{v_0})c_{\pm}, \ \mu_{\pm}=\ln{c_{\pm}}\pm \varphi;\label{1.32}\\
&\varepsilon^2\nabla^2\varphi=N(x)+c^--c^+;\label{1.33}\\
&c_\pm(x,y,0)=c^0_{pm}(y);\label{1.34}\\
&\mathbf{\nabla}^2 \mathbf{v}-\mathbf{\nabla}p+\mathbf{\nabla}^2\varphi\mathbf{\nabla}\varphi=0,\ \mathbf{\nabla}\cdot\mathbf{v}=0;\label{1.35}\\
&c_\pm(x,-L-1,t)=c_\pm(x,L,t)=1,\ \varphi(x,-L-1,t)=-V,\ \varphi(x,L,t)=0.\label{1.36}\
\end{align}
Here $c^0_{\pm}(y),\ \varphi^0(y)$ are 1D steady-state solutions to the problem (\ref{1.31}--\ref{1.36}). Let us consider the instantaneous velocity response, $W$,  in the EBL$\cap$QEB to this perturbation defined as
\begin{equation}
W=\frac{u_t(x,y,0+)}{u_0(x,y)}.\label{1.37}
\end{equation}
To evaluate $W$ we seek a solution to the problem (\ref{1.31}--\ref{1.36}) in the following form:
\begin{equation}
c_\pm=c^0_\pm+tc^1_\pm(y)\exp(iKx),\ \varphi=\varphi^0+t\varphi^1(y)\exp(iKx),\ \mathbf{v}=t\mathbf{v^1}(y)\exp(iKx) \label{1.38}
\end{equation}
Keeping the leading order terms in (\ref{1.31}--\ref{1.36}) we find
\begin{align}
&c^1_\pm+\text{Pe}\left(y(y-1)\right)^2c^0_{\pm y}=0,\ 0<y<L;\ c^1_\pm=0,\ -L-1<y<0;\label{1.39}\\
&\varepsilon^2(\varphi^1_{yy}-K^2\varphi^1)=c^1_--c^1_+,\ -L-1<y<L;\label{1.40}\\
&u^1=(\varphi^0(0)-\varphi^0(y))\Phi^1 iK+A_{C}C^1iK+A_{J}J^1iK+\nonumber\\
&\left(A_{k}-\frac{(\varphi^0(y)-\varphi^0(0))^2}{2k^0}-\frac{\varphi^0(y)-\varphi^0(0)}{k^0}(\varphi^0(y)-\Phi^0+2k^0)\right)k^1iK;
\label{1.41}\
\end{align}
Here
\begin{align}
& C^1=c^1_+=c^1_-;\ J^1=2C^1_{+y};\ I^1=2(C^1\varphi^0_y+C^0\varphi^1_y);\nonumber\\
& k^1=\frac{I^1}{J}-\frac{J^1I^0}{J^{02}};\ \Phi^1=\varphi^1-k^1\ln\left(y+\frac{C^0}{J^0}\right)-k^0\left(\frac{C^1}{J^0}-\frac{C^0J^1}{J^{02}}\right).\label{1.42}
\end{align}
Substituting (\ref{1.30}), (\ref{1.41}) in (\ref{1.37}) we find the following representation of $W$ valid  in the EBL$\cap$QEB:
\begin{align}
&W=\frac{(\varphi^0(0)-\varphi^0(y))\Phi^1}{2y(y-1)(2y-1)}K^2+\frac{A_{C}C^1}{2y(y-1)(2y-1)}K^2+\frac{A_{J}J^1}{2y(y-1)(2y-1)}K^2+\nonumber\\
& \nonumber\\
&\frac{\left(A_{k}-\frac{(\varphi^0(y)-\varphi^0(0))^2}{2k^0}-\frac{\varphi^0(y)-\varphi^0(0)}{k^0}(\varphi^0(y)-\Phi^0+2k^0)\right)k^1}{2y(y-1)(2y-1)}K^2=\nonumber\\
&W_{\Phi}+W_{C}+W_J+W_k;\label{1.42}\
\end{align}

\begin{figure}[!htb]
\includegraphics[width=\columnwidth,keepaspectratio=true]{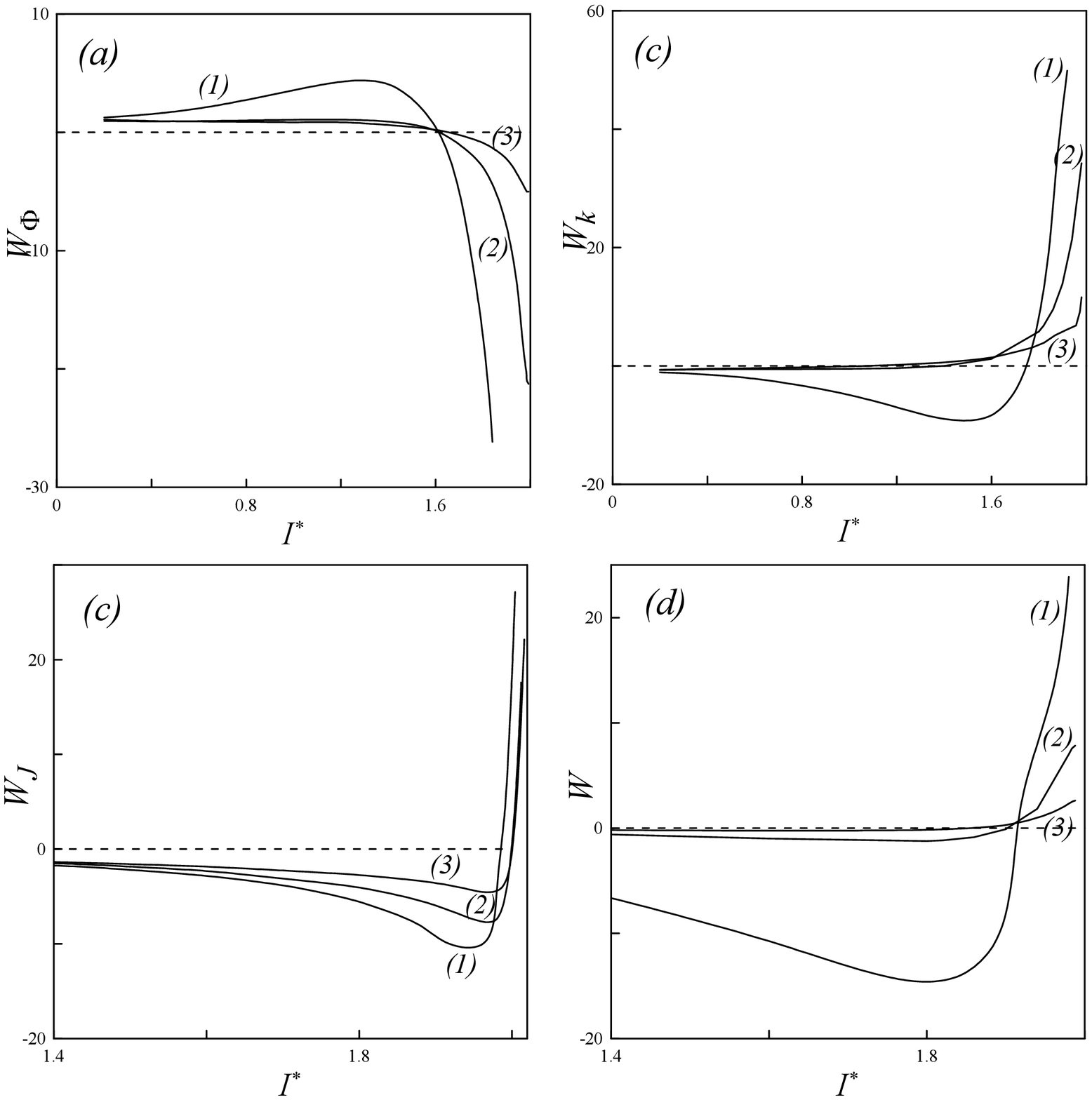}
\caption{Dependence of the contribution of driving factors $W_{\Phi}, W_{J}, W_k$ to the instantaneous velocity response $W$ on the normalised current $I^*\ (I^*=2$  -- the limiting current), $y = 0.05$,  $\varepsilon=10^{-4},\ N=1/3(1),\ 1(2),\ 3(3)$, Pe$=0.5$.  (a) $W_{\Phi}$ versus $I^*$; (b) $W_k$ versus $I^*$; (c) $W_{J}$ versus $I^*$; (d) Total velocity response $W$ versus $I^*$.}
\end{figure}

Here, $W_{\Phi},\ W_{C},\ W_J$ and $W_k$ stand, respectively, for each of the four contributions to $W$. Numerical solution of (\ref{1.31})--(\ref{1.36}) shows that for the entire range of the applied voltage $V$ and membrane's fixed charge density $N$ the contribution of the $W_C$ factor to the instantaneous velocity response $W$ is negative (stabilizing) and negligible compared to that of other terms. The same holds true for the contribution of the electroneutral bulk away from the interface, $y{\not\in}$EBL.  In Fig. 6 we illustrate the contribution of each of the remaining three driving factors $W_{\Phi},\ W_J,\ W_k$ to $W$ (a positive value of a factor corresponds to its destabilising effect through contributing to a positive feedback). We observe that, whereas near the limiting current the contribution of EO factor $W_{\Phi}$ is negative for the entire range of membrane perm-selectivity, Fig.6a, the contribution of its singular bulk analog, $W_k$,  is positive and dominates the hydrodynamic instability for weakly-charged membranes. With the increase of the membrane perm-selectivity the contribution of $W_k$ decreases and the total response, $W$, and so the EC instability are mediated by the non-equilibrium EO factor $W_J$, Fig.6c.

\section{Summary}
In our study we attempt at systematically analyzing the effect of the interface perm-selectivity upon ECI in CP. This includes the analysis of transition from the regime of Non-Equilibrium EO Instability induced by the ESC to the recently discovered equilibrium EO and bulk ECI upon the reduction of the membrane fixed charge density or upon increasing the bulk electrolyte concentration. The goal is to develop a unified description of ECI in CP, valid for the entire range of the membrane perm-selectivity and the applied voltage.

We identify four major factors affecting EC in CP and analyze their effect on the hydrodynamic stability of the system:
\newline 1. Tangential variation of the electrolyte concentration (conductivity, induced space charge, etc). This factor, $A_{C}C_x$, in (\ref{1.29a}) is identified as generalized diffusio-osmosis.
\newline 2. Tangential regular (applied) electric field acting upon the space charge of quasi-equilibrium EDL or the residual space charge of the quasi-electroneutral bulk. This factor,  $(\varphi(x,0,t)-\varphi(x,y,t))\Phi_x$, (\ref{1.29a}), is identified as regular ICEO, \cite{31}, and its bulk analog.
\newline 3. Tangential variation of the counter-ionic transport number in the system responsible for tangential variation of the logarithmic potential drop normal to the interface, forming in the bulk solution in the vicinity of the interface in the course of CP. This factor, $\left(A_{k}-\frac{(\varphi(x,y,t)-\varphi(x,0,t))^2}{2k}-\frac{\varphi(x,y,t)-\varphi(x,0,t)}{k}(\varphi(x,y,t)-\Phi(x,t)+2k)\right)k_x$ in (\ref{1.29a}), constitutes a singular bulk analog of  ICEO.
\newline 4. Tangential variation of the total normal ionic mass flux. This factor, $A_{J}J_x$ in (\ref{1.29a}), may be vaguely termed fricto-osmosis and is entirely related to the extended space charge of the non-equilibrium EDL forming at and above the limiting current.

The contribution of each of these four factors to the hydrodynamic stability of the 1D quiescent steady state conduction may be summarized as follows. For the entire range of the applied voltage $V$ and membrane fixed charge density $N$ the contribution of the first (diffusio-osmosis) factor is slightly stabilizing but generally negligible compared to that of other factors. The second factor (ICEO), while weakly destabilizing away from the limiting current, turns stabilizing near the limiting current for entire range of membrane perm-selectivity; whereas its singular bulk counterpart (the third factor) is destabilizing and dominates the hydrodynamic instability for weakly-charged membranes. With the increase of the membrane perm-selectivity, the contribution of this factor decreases and the overall velocity response and hydrodynamic instability are mediated by the destabilizing forth factor (non-equilibrium EO or fricto-osmosis).


\begin{thebibliography}{100}
\bibitem[1]{1} O.R. Brown and H.R. Thirsk, Electrochim. Acta \textbf{10}, 383 (1965).
\bibitem[2]{2} Y.C. Wang,  A.L. Stevens and J. Han,  Anal. Chem. \textbf{77}, 44293 (2005).
\bibitem[3]{3} M. Rosso, Electrochim. Acta \textbf{53}, 250 (2007).
\bibitem[4]{4} R.B. Schoch, J. Han and P. Renaud, Rev. Mod. Phys. \textbf{80}, 839 (2008).
\bibitem[5]{5} A. Mani, T.A. Zangle and J.G. Santiago, Langmuir \textbf{25}, 3898 (2009).
\bibitem[6]{6} T.A. Zangle, A. Mani and J.G. Santiago, Langmuir \textbf{25}, 3909 (2009).
\bibitem[7]{7} V. Nikonenko, N. Pismenskaya, E. Belova, P. Sistat,  P. Huguet, G. Pourcelly and C. Larchet, Adv. Coll. Interface Sci. \textbf{160}, 101 (2010).
\bibitem[8]{8} S.J. Kim, Y.A. Song and J. Han, Chem. Soc. Rev. \textbf{39}, 912 (2010).
\bibitem[9]{9}  S.J. Kim, S.H.  Ko, K.H. Kang and J.  Han, Nat. Nanotechnol. \textbf{5}, 5 (2010).
\bibitem[10]{10}  T.A. Zangle, A. Mani and J.G. Santiago,  Chem. Soc. Rev. \textbf{39}, 1014 (2010).
\bibitem[11]{11} A. Mani and M.Z. Bazant, Phys. Rev. E \textbf{84}, 061504 (2011).
\bibitem[12]{12} M.B. Andersen, M. van Soestbergen, A. Mani, H.  Bruus, P.M. Biesheuvel and M.Z. Bazant,  Phys. Rev. Lett. \textbf{109}, 108301 (2012).
\bibitem[13]{13}  A. Yaroshchuk, Adv. Colloid Interface Sci. \textbf{68}, 183 (2012).
\bibitem[14]{14} A. Yaroshchuk, Microfluid. Nanofluid. \textbf{12}, 615 (2012).
\bibitem[15]{15}  M. E. Suss, T. F. Baumann, W. L. Bourcier, C. M. Spadaccini, K. A. Rose,  J. G. Santiago and M. Stadermann, Energy and Env. Sc. \textbf{5}, 9511 (2012).
\bibitem[16]{16} 16. H.C. Chang, G. Yossifon and E. A. Demekhin, Ann. Rev. Fluid Mech. \textbf{44}, 401 (2012).
\bibitem[17]{17} D. S. Deng, E.V.Dydek, J.H.Han,S. Schlumpberger, A.Mani, B.Zaltzman and M.Z. Bazant,  Langmuir \textbf{29}, 16167 (2013).
\bibitem[18]{18} R. Kwak, G. Guan, W. K. Peng and J. Han, Desalination \textbf{308}, 138 (2013).
\bibitem[19]{19} R. Kwak, V.S. Pham, K.M. Lim and J. Han, Phys. Rev.Lett. \textbf{110}, 114501 (2013).
\bibitem[20]{20} V. V. Nikonenko, A. V. Kovalenko, M. K. Urtenov, N. D. Pismenskaya, J. Han, P. Sistat and G. Pourcelly, Desalination \textbf{342}, 85 (2014).
\bibitem[21]{21} D. S. Deng, A. Wassim, S. Schlumpberger, M.E. Suss, M.Z. Bazant and W.A. Braff, Desalination \textbf{357}, 77 (2015).
\bibitem[22]{22}	G. Yossifon and H.C. Chang,  PRL \textbf{101}, 254501 (2008).
\bibitem[23]{23}	S.M. Rubinstein, G. Manukyan, A. Staicu, I. Rubinstein, B. Zaltzman, R.G.H. Lammertink, F. Mugele and M. Wessling, Phys. Rev. Lett. \textbf{101}, 236101 (2008).
\bibitem[24]{24} I. Rubinstein and B. Zaltzman, Phys. Rev. E \textbf{62}, 2238 (2000).
\bibitem[25]{25} I. Rubinstein and B. Zaltzman, Math. Mod. Meth. Appl. Sc. \textbf{11}, 263 (2001).
\bibitem[26]{26}  I. Rubinstein, B.  Zaltzman, J. Pretz and C. Linder, Russ. Electrochem. \textbf{38}, 853 (2002).
\bibitem[27]{27}	E.V. Dydek, B. Zaltzman, I. Rubinstein, D.S. Deng, A. Mani and M.Z. Bazant, Phys. Rev. Lett. \textbf{107}, 118301 (2011).
\bibitem[28]{28} V. S. Pham, Z. Li, K. M. Lim, J. K. White and J. Han, Phys. Rev. E \textbf{86}, 046310 (2012).
\bibitem[29]{29} V.G. Levich, \textsl{Physicochemical Hydrodynamics.} Prentice-Hall 1962.
\bibitem[30]{30}	S.S. Dukhin and B.V. Derjaguin, \textsl{Electrophoresis}, Nauka, Moscow, 1976.
\bibitem[31]{31} T. M. Squires and M. Z. Bazant, J. Fluid Mech. \textbf{509}, 217 (2004).
\bibitem[32]{32} S.S. Dukhin, Adv. Colloid Interface Sci. \textbf{35}, 173 (1991).
\bibitem[33]{33}	B. Zaltzman and I. Rubinstein, J. Fluid Mech. \textbf{173}, 579 (2007).
\bibitem[34]{34} B.M. Grafov and A.A. Chernenko, Dokl. Akad. Nauk SSSR \textbf{146}, 135 (1962) (in Russian).
\bibitem[35]{35}  R.P. Buck, J. Electroanal. Chem. \textbf{46}, 1 (1973).
\bibitem[36]{36} A.V. Listovnichy, Elektrokhimia 25, 1651 (1989) (in Russian).
\bibitem[37]{37} V.V. Nikonenko, V.I. Zabolotsky and N.P. Gnusin, Sov. Electrochem. \textbf{25}, 262 (1989).
\bibitem[38]{38} K.T. Chu and M.Z. Bazant, SIAM J. Appl. Math. \textbf{65}, 1485 (2005).
\bibitem[39]{39} W.H. Smyrl and J. Newman,  Trans. Farday Soc. \textbf{63}, 207 (1967).
\bibitem[40]{40} I. Rubinstein and L. Shtilman, J. Chem. Soc. Faraday Trans. II \textbf{75}, 231 (1979).
\bibitem[41]{41}	E.K. Zholkovskij, M.A. Vorotynsev and E. Staude, J.  Coll. Int. Sc. \textbf{181}, 28 (1996).
\bibitem[42]{42} I. Rubinstein and B. Zaltzman, Phys. Rev. Lett. \textbf{114}, 114502 (2015).
\bibitem[43]{43}	I. Rubinstein and B. Zaltzman, in \textsl{Surface Chemistry and Electrochemistry of membranes}, T.S. Sorensen (ed.) Marcel Dekker, Inc., New York (1999).
\bibitem[44]{44} C.L. Druzgalski, M. B. Andersen and A. Mani,  Phys. Fluids \textbf{25}, 110804 (2013).
\bibitem[45]{45}	H.C. Chang, E. A. Demekhin and V. S. Shelistov, Phys. Rev. E \textbf{86}, 046319 (2012).
\bibitem[46]{46} E. A. Demekhin, N. V. Nikitin and V. S. Shelistov,  Phys. Fluids \textbf{25}, 122001 (2013).
\bibitem[47]{47} A.S. Khair., Phys. Fluids \textbf{23}, 072003 (2011).
\bibitem[48]{48} E.A.Demekhin, S. Amiroudine, G.S. Ganchenko and N.Yu. Khasmatulina, Phys. Rev. E \textbf{91}, 063006 (2015).
\bibitem[49]{49} I. Rubinstein and B. Zaltzman, Phys. Rev. E \textbf{81}, 061502 (2010).

\end{thebibliography}
\end{document}